# *DefTesPY*: Cyber defense model with enhanced data modeling and analysis for Tesla company via Python Language

Naresh Kshetri, CyROC, Emporia State University, Emporia, Kansas, USA, NKshetri@emporia.edu
Irin Sultana, CyROC, Emporia State University, Emporia, Kansas, USA, ISultana@g.emporia.edu
Mir Mehedi Rahman, CyROC, Emporia State University, Emporia, Kansas, USA, MRahman2@emporia.edu
Darshana Shah, CyROC, Emporia State University, Emporia, Kansas, USA, DShah@emporia.edu

**Abstract** - Several types of cyber-attacks on automobiles and business firms keep on rising as we are preparing to counter cybercrimes with several new technologies and defense models. Cyber defense (also, counter intelligence) is a computer network defense mechanism that involves response to activities, critical infrastructure protection, and information assurance for corporations, government bodies, and other conceivable networks. Cyber defense focuses on preventing, detecting, and responding to assaults or threats in a timely manner so that no infrastructure or information is compromised. With the increasing volume and complexity of cyber threats, most companies need cyber defense to protect sensitive information and assets. We can control attacker actions by utilizing firewalls at different levels, an intrusion detection system (IDS), with the intrusion prevention system (IPS) which can be installed independently or in combination with other protection approaches. Tesla is an American clean energy and automotive company in Austin, Texas, USA. The recent data breach at Tesla affected over 75,000 individuals as the company pinpoints two former employees as the offender revealing more than 23,000 internal files from 2015 to 2022. In this work, we will emphasize data modeling and data analysis using cyber defense model and python with a survey of the Tesla company. We have proposed a defense model, DefTesPY, with enhanced data modeling and data analysis based on the encountered cyber-attacks and cybercrimes for Tesla company till date.

## I. INTRODUCTION

Tesla is already a globally recognized producer and distributor of electric vehicles. It continues to promote and lead the growth of the electric vehicle sector, updating people's understanding of the technology. Tesla has been said to be a company that has the power to transform the planet [1]. In today's digital world, cybersecurity is more crucial than ever, particularly for forward-thinking organizations like Tesla, which is known for innovating sustainable energy and automotive technology. However, an amazing invention comes with huge risk, as Tesla realized during a recent data breach that compromised the personal information of tens of thousands of people. Tesla's most recent data compromise resulted in the disclosure of over 23,000 internal files from 2015 to 2022, affecting more than 75,000 individuals. The company has identified two former employees as the perpetrators [2] [3]. This event underline[1]s the crucial need for stronger cybersecurity solutions that can adapt to the various strategies used by modern cyberattacks.

---

The dynamic and ever-changing nature of cyberattacks, such as cryptojacking and fileless ransomware targeting the financial sector, which is very dynamic in nature, underscores the importance of the automobile industry remaining vigilant and prepared to counter fresh and emerging threats. The study investigates the tactics used by attackers to keep their malware inactive and frequently changes its variants in order to avoid detection . These attackers strategically wait for a long time, remaining dormant until they launch their attacks on the correct device at the right time to cause the most damage . As a result, the vehicle industry must exercise caution when dealing with potential threats that may exploit flaws in automotive systems [4].

Cybersecurity cannot be established or maintained with a single, silver bullet solution. To deal with sophisticated attacks, a combination of defense techniques that may collaborate and complement one another is required [5]. Cyber defense, often known as counter-intelligence, is a complete approach that includes responding to actions, protecting key infrastructure, and ensuring information security for enterprises, government agencies, and other possible networks. It highlights the importance of anticipating, recognizing, and responding quickly to cyberattacks in order to maintain the integrity of infrastructure and information.

DefTesPY employs a multi-layered defense strategy that includes firewalls, intrusion detection systems (IDS), and intrusion prevention systems (IPS), either as stand-alone solutions or in conjunction with other security measures and proposes a model. These technologies act as the first line of defense against unwanted access, monitoring network traffic for suspicious activity and taking preemptive actions to intercept potential threats. DefTesPY is more than just a response to Tesla's recent data leak; it is a proactive strategy to redefine cybersecurity safeguards for the automobile industry. DefTesPY attempts to provide a dynamic defensive strategy that can respond to new cyber threats while protecting sensitive information and assets from unwanted access and potential breaches.

This study will highlight the capabilities of DefTesPY in improving Tesla's cyber security strategy by focusing on the cyber-attacks and cybercrimes that have occurred to date. We propose an improved and comprehensive defense model that solves existing weaknesses while also anticipating future threats based on a thorough analysis and survey of Tesla's present cybersecurity measures. In doing so, DefTesPY demonstrates the fundamental necessity of cybersecurity in the age of digital transformation, providing useful insights and solutions that other firms may use to protect themselves from the ever-increasing tide of cyber-attacks.

## II. RELATED WORK

In [6], Goodall and Sowul (2009) identified visual analytics technologies as a crucial aspect in the field of cyber defense to counter the escalating intricacy and quantity of cyber threats. VIAssist is a visual analytics system that aims to improve situational awareness for cyber-infrastructure defenders, which is a significant advancement in this industry. VIAssist overcomes the shortcomings of conventional tools such as spreadsheets and fundamental charting, which do not utilize the human perceptual capabilities to identify patterns and irregularities in extensive datasets. VIAssist facilitates a more thorough investigation of cyber incidents by incorporating visual exploration, IP geo-location, scalability, collaboration, and reporting functionalities. It enables many degrees of visual analysis, ranging from broad overviews to specific textual data, which aids in comprehending multidimensional information that are crucial for safeguarding vital infrastructure. VIAssist includes a configurable dashboard and synchronized multiple views, connecting distinct visualizations to offer diverse insights into data. This system prioritizes scalability by incorporating features such as Smart Aggregation, which ensures the ability to handle and analyze large datasets while maintaining data manageability and situational awareness. These developments demonstrate the significance of advanced visual analytics in contemporary cyber security methods, which are in line with the objectives of the DefTesPY model suggested for Tesla to enhance data modeling and analysis using Python.

Cyber-cognitive situation awareness (CCSA) is crucial in the realm of cyber defense since it enables human defenders to make successful decisions. In [7], Gutzwiller et al. (2016) highlight the significance of human-centered awareness in cyber defense, acknowledging that although data fusion and automated technologies are valuable, they cannot completely substitute the nuanced comprehension that human analysts contribute to cybersecurity. Their research entails doing a cognitive task analysis (CTA) with the objective of finding crucial pieces of information that analysts need to observe in order to retain situational awareness. The findings emphasize three primary areas of defense awareness: network infrastructure and status, network behavior, and customer actions. Analysts must possess a comprehensive understanding of both typical and atypical patterns inside their networks, grasp the condition and interconnections of network components, and remain knowledgeable about any activity that could potentially affect network security. This thorough comprehension aids in accurately identifying and reducing cyber dangers. The study highlights the importance of incorporating cyber-cognitive situation awareness measures into cyber security operations. It suggests that tools and methods should be developed to meet the cognitive requirements of analysts, hence improving their capacity to safeguard critical infrastructure.

In [8], Zhong et al. (2017) examine the complex data sorting tasks carried out by cyber defense analysts in Security Operations Centers (SOCs), highlighting the crucial importance of cognitive processes in cyber situational awareness (Cyber SA). Their research presents an advanced human-machine system that aims to enhance data triage by recording and analyzing the cognitive patterns of experienced analysts. The cognitive traces represent the decision-making pathways and analytical techniques used by expert analysts. They form a context-based retrieval system that allows for the immediate retrieval of relevant cognitive traces based on the current analytical context. The system integrates network monitoring data, observed attack activities, and the growing mental models of analysts using a detailed cognitive task analysis (CTA) framework. This framework carefully records the actions, observations, and hypotheses (AOH) of analysts. The data is organized using AOH-trees and hypothesis-trees (H-trees) to illustrate the cognitive processes used to identify and respond to potential cyber threats. The system employs cognitive traces to mitigate the problem of a high noise-to-signal ratio in network monitoring data. It filters away extraneous data and emphasizes significant events, hence simplifying the triage process. The implementation of this system showcased enhanced precision and swiftness in identifying and addressing threats, highlighting the significance of combining human cognitive analysis with modern automated technologies to bolster the overall effectiveness of cyber defense operations.

In [9], Freitas De Araujo-Filho et al. (2021) conducted research in the field of cybersecurity for automotive systems. They focused on developing an intrusion prevention system (IPS) for Controller Area Networks (CAN) that is very effective. Their work is closely related to the aims of our DefTesPY model. Their Intrusion Prevention System (IPS) uses machine learning to identify and impede cyber-attacks on Controller Area Network (CAN) without modifying the Electronic Control Units (ECUs) or necessitating manufacturer-specific data. The system uses unsupervised techniques such as Isolation Forest and One-Class Support Vector Machine (OCSVM) to detect abnormalities by identifying deviations from regular CAN traffic patterns. This method attains a notable level of precision and swift identification durations, which are crucial in order to avert harm caused by malevolent frames. The DefTesPY model employs sophisticated data modeling and machine learning techniques in Python to improve cyber defense mechanisms for Tesla's network. It specifically emphasizes prompt threat identification and response. Both methods emphasize the urgent requirement for strong, quick intrusion detection systems that can function on affordable technology, guaranteeing wide usability and fast implementation in the automotive industry.

In [10], Malik and Sun (2020) demonstrated a comprehensive analysis and simulation of different cyber-attacks on connected and autonomous vehicles in the cybersecurity arena. Their research is highly relevant to the DefTesPY model. Their study brought attention to the growing intricacy and escalating cyber-security risks linked to smart connected cars, underscoring the importance for users to comprehend the security ramifications prior to operating these vehicles. Through the use of threat modeling and assault simulations, they successfully discovered and showcased the consequences of major cyber threats such as ransomware, remote control takeovers, and network vulnerabilities. Their findings emphasize the possibility of severe consequences, including extensive data breaches and physical harm to users. These coincide with the commitment to improving data modeling and cyber security systems in Tesla's network. The simulations conducted by the researchers offer valuable knowledge and approaches that enhance the advanced data modeling and machine learning methods used in DefTesPY. The goal is to strengthen the cybersecurity of automotive systems by protecting them against complex cyber threats.

**III. CYBER ATTACKS ON AUTOMOBILES**

The cyber-attacks against next generation cars (say, electric cars, hybrid cars, self-driving cars) along with automobiles, including connected and autonomous vehicles is increasing. Vehicle manufacturers are facing unique levels of challenge that increase complexity, as modern electric vehicles (EVs) are becoming more vulnerable to cybersecurity threats [11] [12]. More vehicles sold in the United States are connected vehicles, that is , they are already vulnerable to cyber-attacks. There can be the loss of human lives in case of cyber-attacks against automobiles. The use of machine learning and deep neural networks (DNN) against cyber-attacks for automated vehicles can provide effective cybersecurity for tracking status and data visualization [13].

When it comes to vehicle/vehicular networks, hackers are accessing all vehicle-to-vehicle (V2V), vehicle-to-pedestrian (V2P), and vehicle-to-infrastructure (V2I) networks. They can easily broadcast false messages on smart systems of vehicles and use malwares to break the security system in networks and connected vehicles [14]. When it comes to self-driving cars, the situation is even worse and prone to cyber-attacks. Technological advancements in connected vehicles to make real self-driving cars (or even as driverless cars) as improved efficiency, road safety, traffic management, and driver comfort with safe means of transport is the loophole and entry point for malicious cyber-attacks in many connected autonomous vehicles (CAVs) [15]. Some of the recent attacks on self-driving cars are malware attack, man-in-the-middle (MITM) attack, denial of service (DOS) attack, ransomware attacks, spoofing attacks, sybil attacks, illusion attacks, and impersonation attacks.

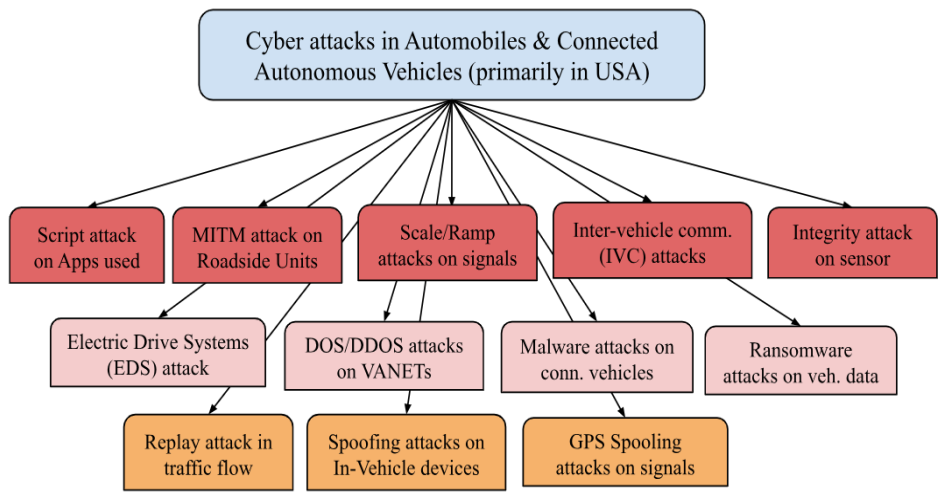

Figure 1: Types of cyber-attacks in automobiles and connected autonomous vehicles (CAVs) primarily in the United States (US) [6]-[10].

## IV. CASE STUDY FROM TESLA

Tesla was founded in 2003 by a group of engineers who aimed to show that driving an electric vehicle does not require sacrifice. The company produces not only all electric vehicles, but also sustainable energy producing and storage technologies. Tesla advocates for reducing reliance on fossil fuels and achieving zero carbon emissions. [16][19]. It has quickly become one of the world's leading manufacturers of electric vehicles. This success can be attributed, at least in part, to their numerous technological inventions, which range from batteries to electric drives [20]. In fact, Tesla became the market leader in the battery sector by recognizing its potential early on and perfecting its R&D technology. Tesla is the only electric vehicle company that uses a three-yuan 18,650 lithium-ion battery. The 18650-lithium-ion ternary battery offers great energy density, stability, and consistency. [14]. Another feature of Tesla Superchargers is the world's most advanced charging technology, taking only 20 minutes to fill half of the battery and charging far faster than other charging stations [17].

Tesla chose an induction motor for its Model S, the world's best-selling electric car in 2015 and 2016, despite its lower efficiency and power-to-weight ratio compared to the more commonly used permanent magnet motors in the automotive industry. However, the Model S still delivers remarkable performance in acceleration, power output, and range, sparking engineers' interest in its design and conception. However, Tesla patents provide key geometric dimensions for modeling, using both specific values and ratios. Selected dimensions ensure model performance, aiming for practical accuracy for educational and engineering purposes, with highly uncertain data highlighted in italics in the reference table [20].

| Parameter name | Symbols used | Values used in the proposed model |
|---|---|---|
| Stack length | L | 152 mm |
| Air gap | e | 0.5 mm |
| **Stator** | | |
| Stator outer diameter | $D_{OS}$ | 254 mm |
| Stator inner diameter | $D_{IS}$ | 156.8 mm |
| Slot depth | $L_{sd}$ | 19 mm |
| Slot opening width | $W_{so}$ | 2.9 mm |
| Tooth tip depth | $L_{td}$ | 1 mm |
| Tooth width | $W_{ts}$ | 4.5 mm |
| **Rotor** | | |
| Rotor outer diameter | $D_{OR}$ | 155.8 mm |
| Rotor inner diameter | $D_{IR}$ | 50 mm |
| Slot depth | $L_{rd}$ | 19.6 mm |
| Slot bridge | $L_{br}$ | 0.55 mm |
| Tooth width | $W_{tr}$ | 4.5 mm |

Figure 2: Tesla Model S (best-selling electric car in 2015 and 2016) Induction motor main dimensions [20]

In addition, Tesla unveiled their humanoid robot prototype, "Optimus," on September 30th, 2021. It has extensive experience with artificial intelligence. Elon Musk claims that the bot will use an optical neural network to manage tasks autonomously. Tesla's autonomous driving solution relies on a perception network that generates a three-dimensional space from the real world. A hybrid planning system combines traditional methods with neural algorithms to plan the car's behavior and trajectory. However, commercialization of the robot is slow due to its lack of practical solutions and high cost [18].

Moreover, despite government incentives, the Model 3 remains beyond see of reach for many drivers, costing $43,990 in tax credits and gas savings as of May 2021. To be competitive in the new energy vehicle business, Tesla must discover ways to reduce expenses and lower product prices. The Chevy Bolt and Nissan Leaf, two well-known competitors, struggled to gain early traction because of high costs and limited range [19].

**V. NEED FOR CYBER DEFENSE AND INTELLIGENCE**

Cybersecurity and the information security of vehicular networks are primary needs when it comes to automobiles safety and road safety. Securing the vehicular network against cyber attackers is a game between two players - leader agent and follower agent [21]. Malicious users are able to implement various kinds of attacks when it comes to connected autonomous vehicles. In-vehicle network attacks are based on types of communication networks and attack objects that classifies several cybersecurity vulnerabilities and risks [22]. The use of sensors in modern vehicles are another cybersecurity risks and primary cybersecurity threats. There are primary two sensor types as Vehicle dynamic sensors and Environment sensors that can compromise the security of the control layer in connected vehicles [23]. Through advanced vehicular sensors, all the autonomous vehicles (including Tesla connected cars) identify road signs, recognize collision threats, estimate distance between objects and vehicles, and monitor roadways.

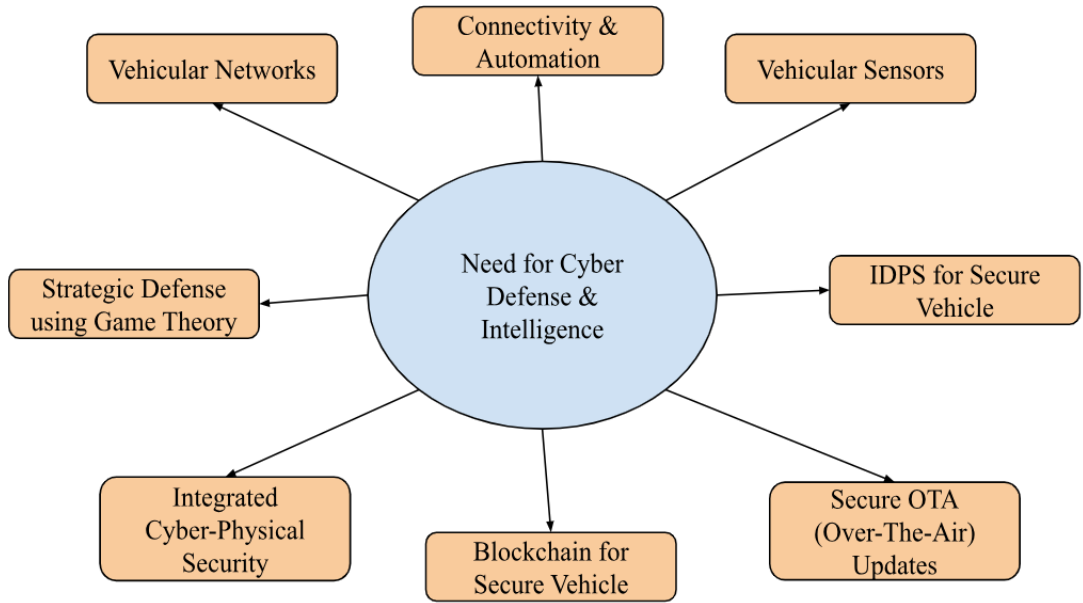

Figure 3: The need for cyber defense and counter intelligence for automobiles and connected vehicles including Tesla cars [21] [22] [23]

The growing frequency of cyber-attacks on vehicle networks highlights the necessity for strong cyber protection and intelligence in these systems. A complete strategy for cyber defense, such as the Stackelberg game-based scheme, involves the use of numerous security agents, including IDS, IPS, and IRS. These agents are coordinated by an Intrusion Decision Agent (IDA) to reduce costs and enhance security efficacy [21]. This hierarchical strategy guarantees the optimal operation of each component and utilizes game theory to anticipate and minimize attacks before substantial harm takes place, thereby dealing with the excessive costs and inaccuracies often linked to current protection mechanisms [21]. Advanced detection, prevention, and response tactics are necessary to protect critical infrastructure in connected and autonomous cars (CAVs) from cyber-security concerns, such as assaults on in-vehicle systems and V2X networks [22]. The proposed defensive model, DefTesPY, utilizes advanced data modeling and analysis to tackle specific issues encountered by modern vehicular networks. This approach aims to improve resilience against emerging cyber-attacks [22]. Incorporating advanced technologies such as machine learning and artificial intelligence enables the prompt identification of abnormalities and potential dangers in real-time. Furthermore, the use of blockchain technology enhances the security of communication and transaction procedures by offering an unchangeable and transparent record [23]. Tesla may greatly improve the robustness of their vehicle networks by implementing modern approaches. This will ensure the safety and reliability of their systems, as well as protect sensitive information and vital infrastructure from increasing cyber threats [22][23].

**VI. DEFENSE MODEL - *DefTesPY* Proposed Model**

The need for a cyber defense model at this moment is urgent and alarming as cyber-attacks in automobiles and connected vehicles (including Tesla and other driverless car manufacturing companies) are increasing. The proposed DefTesPY model is shown below in Figure 3. The components of the cyber defense model (Users, Monitoring Systems, Vehicular Networks, Tesla Intrusion Prevention System, Tesla Intrusion Detection System, Firewalls, Attackers), DefTesPY are outlined and described in the sections below:

*A. Users:* Users in the cyber defense model are the car users (Tesla) including car drivers, car passengers, and also pedestrians (those nearby the operating Tesla cars). The non-living moving objects that appear suddenly on the roads are not identified as users due to some constraints.

*B. Monitoring system(s):* The monitoring and analysis system in the DefTesPY cyber defense model connects with vehicular networks (like V2V, V2I, others etc.) and communicates with the car users.

*C. Vehicular network(s):* Several networks include vehicular networks for cars on the top of Level 3 TIPS in DefTesPY cyber defense model and also the Tesla network that is created and installed for vehicle safety and security.

*D. Tesla IPS:* Intrusion prevention system (IPS) for DefTesPY are the hardware and/or software devices installed on vehicular networks to prevent any identified or detected cyber-attacks in Level 2 by TIDS. The use of software (instead of hardware devices) is quite popular for intrusion prevention systems but is the primary target of cyber-attacks today.

*E. Tesla IDS:* Intrusion detection system (IDS) in Level 2 of DefTesPY cyber defense model are also hardware and/or software devices installed only to detect cyber-attacks. The cyber-attacks for vehicular systems that surpass the initial level 1 (of firewalls) are detected in this level (level 2) of proposed cyber defense model.

*F. Firewalls:* Firewalls are the software applied as Level 1 security as primary perimeter defense or counter measures in our DefTesPY cyber defense model. Attacks bypassed by Level 1 should be detected by Level 2 before passing to Level 3.

*G: Attacker(s):* In the cyber world, we treat every unknown user, penetration tester as possible cyber attackers and hackers who can launch several cyber-attacks on automobiles from GPS spoofing attack on signals to script attacks on apps installed.

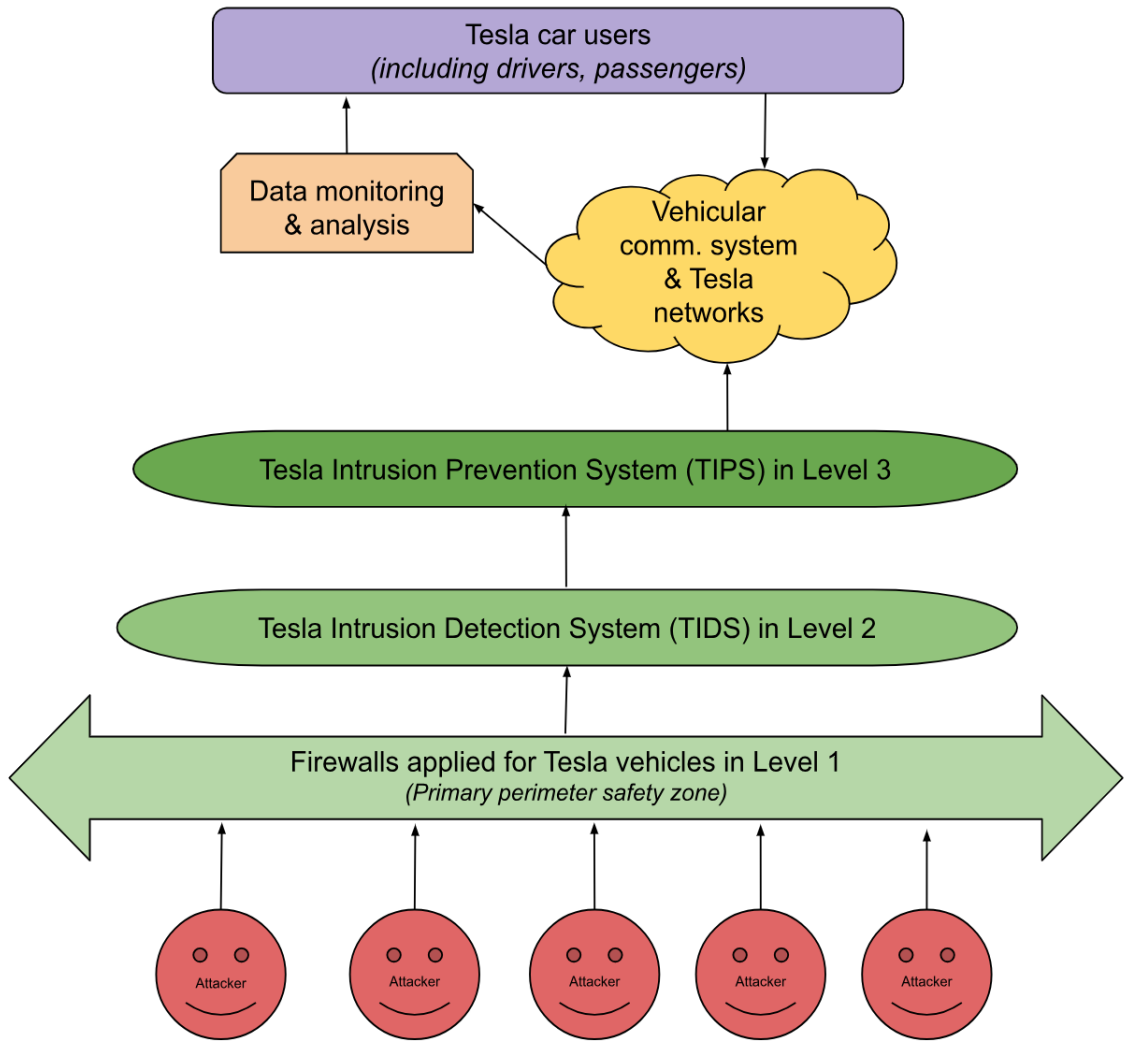

Figure 4: Proposed model - DefTesPY cyber defense model with enhanced data modeling and analysis at different levels based on cyber-attacks and threats for Tesla car

The algorithm for our proposed cyber defense model DefTesPY is given as Algorithm - DefTesPY.

---

**Algorithm - DefTesPY (proposed cyber defense model)**

---

1 tips/tids = calculating probability of Tesla's system intrusion detected and or intrusion prevented
2 lev [n] = selecting a level to apply the defense strategies as of cyber-attacks, n is level number
3 attk = determines the attack version as initial, detection, or prevention for attackers
4 psz = denotes primary perimeter safety zone as indication of firewall
5 def = generating defense strategies as per lev [n] determined to counter the cyber attacks

```
6 while (the cyber-attack on Tesla keeps continuing)
7        psz = select perimeter safety zone via levels
8        if (the attk = initial && tips/tids = 0)
9               def = apply firewall as primary lev [1]
10       else psz = psz + 1
11       end
12 end
```

## VII. ANALYSIS & CONCLUSION

The analysis of current cyberattacks and ongoing malwares injection on connected vehicles sought immediate need of more counterintelligence and cyber defense models in autonomous vehicles in the United States including Tesla. As the United States remains one of the most powerful cyber nations in the world (followed by China, Russia, UK, and Australia) [24], there is no doubt that the critical infrastructure of the United States is the most targeted, including the automobile industry. Vehicular networks are the primary attack base by cyber attackers as the connected automobile industry continues to grow in the future. Attacks on Tesla's self-driving cars due to remote connection features and connected network technologies is still a huge challenge. Self-driving cars can also be used by cyber criminals and nation states terrorists to harm or attack any public places. The use of Autopilot in Tesla cars have reported several crashes not only including drivers and passengers but also harm to pedestrians. The potential risk of being hacked and controlled by cyber hackers, malfunctioning of automated systems in the automation, no response to animals, objects, or people that appear suddenly, risk with health are still primary dangers associated with self-driving cars including Tesla.

As hackers can easily trick self-driving cars and make them run in unwanted ways, the need of cyber defense models like DefTesPY is urgent for every type of cyber-attacks including ransomwares on vehicle data and inter-vehicle communication attacks. In the bestselling list of cars, Tesla Model Y made it to top five, and Tesla Model 3 made it to top thirteen as Toyota Corolla, Ford F-Series, and Volkswagen Golf are the three best selling cars of all time in the United States [25]. Like the purchasing and selling of cars, incidents of cyberattacks continue to rise and remain serious threats to the United States as nation-state hackers (Russia, China) being the biggest cyber threats to all critical infrastructure (including connected vehicles) of the United States.

## VII. FUTURE SCOPE

The future scope of cyber defense and counter intelligence (whether it is in the automobile industry like Tesla or in the healthcare industry) will be needed more for sure. Attackers are deploying several tactics to trick users (like Tesla car users), and inject malwares on security systems. The cyber defense model can be identified, introduced, and proposed for every car company and for data monitoring and data analysis. Analyzing threats on self-driving cars and attacks on intelligent traffic systems (ITS), and their impacts on car passengers (including pedestrians safety) along with vulnerabilities determines the future of smart transportation. We definitely need to improve security issues and ongoing cyber-attacks associated with self-driving cars on the basis of vehicle data communication like vehicle-to-vehicle (V2V), vehicle-to-infrastructure (V2I), and others.